\documentclass[prb,twocolumn,superscriptaddress, reprint]{revtex4-2}

\usepackage{amsmath, amssymb}
\usepackage{graphicx}
\usepackage{float}
\usepackage[colorlinks = true, urlcolor = cyan, linkcolor=blue, citecolor=blue, filecolor=magenta]{hyperref}
\usepackage{adjustbox}
\usepackage{orcidlink}

\newcommand{\ket}[1]{|#1\rangle}

\begin{document}

\title{Experimental Quantum Bernoulli Factories via Bell-Basis Measurements}

\author{Tanay Roy\,\orcidlink{0000-0001-9442-862X}}
\email{roytanay@fnal.gov}
\affiliation{Superconducting Quantum Materials and Systems (SQMS) Center, Fermi National Accelerator Laboratory, Batavia, IL 60510, USA}

\date{\today}

\begin{abstract}
Randomness processing in the Bernoulli factory framework provides a concrete setting in which quantum resources can outperform classical ones. We experimentally demonstrate quantum randomness processing based on Bell-basis measurements of two identical input quoins prepared on IBM superconducting hardware. Using only the measurement outcomes (and no external classical randomness source), we realize the classically inconstructible Bernoulli doubling primitive $f(p)=2p$ and, as intermediate outputs from the same Bell-measurement statistics, an exact fair coin $f(p)=1/2$ and the classically inconstructible function $f(p)=4p(1-p)$. We benchmark the measured output biases against ideal predictions and discuss the impact of device noise. Our results establish a simple resource-efficient experimental primitive for quantum-to-classical randomness processing and support the viability of quantum Bernoulli factories for quantum-enhanced stochastic simulation and sampling tasks.
\end{abstract}

\maketitle

\section{Introduction}
Randomness is a foundational resource in computation, cryptography, and stochastic simulation, yet in practice the available source of randomness is often imperfect. A canonical abstraction is a coin with unknown bias $p$ (a $p$-coin), which can be tossed repeatedly but whose value of $p$ is not known a priori. The Bernoulli factory problem asks whether one can transform such a source into a new coin with bias $f(p)$ using only a finite (but random) number of samples---and crucially without estimating $p$---a question rooted in early randomness-extraction ideas such as the von Neumann scheme~\cite{von1951} and later formalized by Keane and O'Brien~\cite{Keane1994Bernoulli}. These protocols produce either a valid output coin or a null outcome after consuming a finite number of $p$-coin tosses, which are then discarded. Crucially, the procedure must terminate with probability 1 and use only a finite number of $p$-coins on average for generating an $f(p)$-coin. 

For classical Bernoulli factories (CBFs), the set of exactly constructible functions $f(p)$ is severely constrained. In particular, functions that approach $0$ or $1$ too rapidly or that attain the values $\{0,1\}$ in the interior of the domain are not constructible by any algorithm that halts with probability one~\cite{Keane1994Bernoulli}. Even when $f(p)$ is classically constructible, the expected sample complexity can be large and strongly dependent on $p$~\cite{von1951, Nacu2005fast}.

Quantum information processing changes this landscape by allowing the input bias to be encoded coherently in a qubit state (a $p$-quoin) and by exploiting measurements on one or more copies of that state. Such quantum Bernoulli factories (QBFs) can implement transformations $p\mapsto f(p)$ that are impossible or highly inefficient classically~\cite{dale2015provable}. Prior experimental demonstrations using small numbers of qubits have shown feasibility for selected classically inconstructible functions in superconducting and photonic platforms~\cite{yuan2016one_qubit, Patel2019expt, zhan2020experimental}. More recently, a general QBF framework with photonic experiments has been reported~\cite{liu2020generalQBF}, and the model has been extended to quantum-to-quantum Bernoulli factories, where both the input and output are quantum states~\cite{jiang2018qqbf, Hoch2025modular, Rodari2024qdqqbf, Paesani2025processing, Hoch2025complexity}.

Here we report an experimental quantum-to-classical randomness-processing primitive based on a single Bell-basis measurement performed on two identical input quoins. This measurement yields, in one shot, rich statistics to generate multiple useful output coins, enabling a compact and resource-efficient implementation. Specifically, we (i) experimentally realize the celebrated Bernoulli doubling function $f_\wedge(p)$ (a classically inconstructible primitive) without using any external classical coin; (ii) concurrently generate an exact fair coin and the classically inconstructible function $f_\frown(p)=4p(1-p)$ as intermediate outputs from the same Bell-measurement data (two-bit output); (iii) demonstrate constant average quoin cost for the fair-coin and $f_\frown(p)$ constructions (two and four quoins, respectively), independent of $p$.

We implement the protocol on IBM superconducting hardware and benchmark the observed output biases against theory, highlighting both the constant-input-cost features of the Bell-measurement approach and the practical limitations arising from device noise. Our approach is related in spirit to prior demonstrations of quantum Bernoulli factories~\cite{Patel2019expt}, but differs in that the fair coins required for subsequent classical subroutines (e.g., to realize $\sqrt{p}$~\cite{sqrt_p1999}) are generated internally by the same Bell-measurement statistics, avoiding the need for additional external randomness sources.

\section{Fair coin generation}
\label{sec:fair_coin}

\begin{figure}[tb]
    \centering
    \includegraphics[width=\columnwidth]{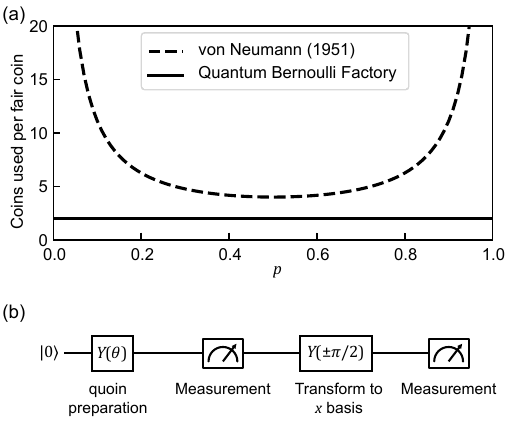}
    \caption{\textbf{Fair coin generation.} (a) The number of biased coins consumed to generate a fair coin using the classical von Neumann method is shown in the dashed line as a function of bias $p$. The solid line shows that a quantum Bernoulli method needs only two quantum coins or quoins per fair coin irrespective of $p$. (b) A quantum circuit using a single qubit for generating a fair coin with $\theta = 2\sin^{-1}\sqrt{p}$ where $Y(\theta)$ represents a rotation about the $y$-axis by an amount $\theta$ and the result of the last measurement is the outcome.}
    \label{fig:fig1}
\end{figure}

Let us first consider the case of generating a fair coin $f(p)=\frac{1}{2}$ using a series of classical coins with unknown bias $p$, called a $p$-coin. Here $p$ denotes the probability of getting heads. Provided $p\neq \{0,1\}$, a solution to this CBF problem was provided by John von Neumann~\cite{von1951}. Toss the $p$-coin twice. If the two outcomes are different (which happens with a probability of $2p(1-p)$) then output the result of the second toss, otherwise discard the two coins and start over. The probability of obtaining heads in this process is 
\begin{equation}
    P(\text{heads}) = \sum_{k=0}^\infty \left[ p^2 + (1-p)^2 \right]^k p(1-p) = \dfrac{1}{2}.
\end{equation}
However, the number of $p$-coins $N_p$ used to generate a fair coin depends on the bias $p$. The probability of generating a coin (heads or tails) is given by $2p(1-p)$ which utilizes two coins, and hence
\begin{equation}
\label{eq:N_p_von}
    N_p = \frac{\text{Number of coins used per round}}{\text{Probability of success per round}} = \frac{1}{p(1-p)}.
\end{equation}
The dashed line in Fig.~\ref{fig:fig1}(a) shows this dependence. While the best case ($p=\frac{1}{2}$) requires 4 coins, $N_p$ diverges when $p$ approaches 0 or 1.

A quantum Bernoulli factory provides a very efficient protocol for $f(p)=\frac{1}{2}$ using either a single qubit or a Bell measurement on two qubits (discussed later). The quantum equivalent of a classical $p$-coin can be encoded as
\begin{equation}
    \ket{p} = \sqrt{1-p}\ket{0} + \sqrt{p}\ket{1},
\end{equation}
where $p\in [0,1]$ and $\sigma_z=\{ \ket{0}, \ket{1} \}$ represents the computational basis corresponding to tails and heads, respectively. We call $\ket{p}$ a $p$-quoin, following the nomenclature in Ref.~\onlinecite{dale2015provable}. When we measure a quoin in any basis, we project it along that basis and obtain a classical answer consuming the quoin. A quoin can show remarkable advantages in randomness processing~\cite{dale2015provable}. As an example, it is evident that a fair coin can be generated by performing only two measurements: first along a random direction and then along a direction orthogonal to it. An example circuit is shown in Fig.~\ref{fig:fig1}(b) where the $p$-quoin is generated by applying the $Y(\theta)$ gate to $\ket{0}$ with $\theta=2\cos^{-1}\sqrt{p}$, which corresponds to a rotation about the $y$-axis. The first measurement is performed along the $z$-axis, and then a measurement is performed along the $x$-axis which involves a basis transformation using a $Y(+\pi/2)$ or $Y(-\pi/2)$ rotation. Thus, a fair coin can be generated using only two quantum measurements or quoins independent of $p$ (see the solid line in Fig.~\ref{fig:fig1}(a)). Notably, this method includes $p=0$ and 1, which are strictly excluded in CBF.

\section{Coin generation using Bell measurements}
\label{sec:Bell_meas}

\begin{figure}[t]
    \centering
    \includegraphics[width=\columnwidth]{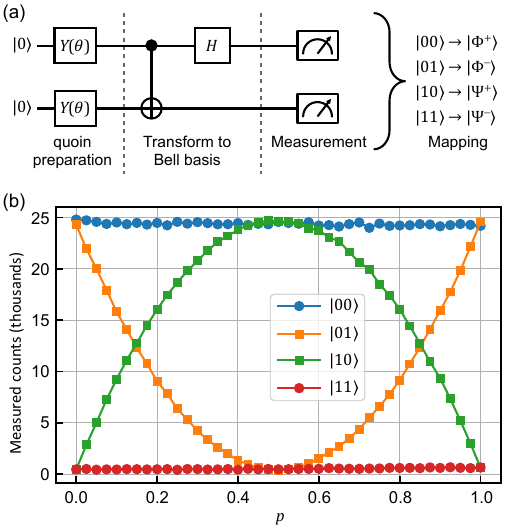}
    \caption{\textbf{Experimental protocol and measurements.} (a) Two quoins are prepared with identical bias $p$ by applying rotations about the y-axis by an amount $\theta=2 \sin^{-1}\sqrt{p}$. A transformation to the Bell-basis is performed by a $\mathtt{CNOT}$ followed by a Hadamard gate. The measured states are mapped to the corresponding Bell basis. (b) Measured counts of the four basis states as a function of bias $p$ obtained after $5\times10^4$ experimental repetitions.}
    \label{fig:fig2}
\end{figure}

The single quoin utilizes quantum coherence for efficient generation of a fair coin. It can also be used to construct other classically disallowed functions, albeit consuming a large number of single quoins~\cite{yuan2016one_qubit}. We now consider utilizing another quantum property, entanglement, for efficient generation of a fair coin together with other functions not allowed in CBF. This entanglement is processed through a Bell measurement on two $p$-quoins $\ket{p}^{\otimes 2}$ where the Bell basis is defined as
\begin{subequations}
\begin{align}
    \ket{\Phi^{\pm}} &= \dfrac{\ket{00}\pm\ket{11}}{\sqrt{2}}, \\
    \ket{\Psi^{\pm}} &= \dfrac{\ket{01}\pm\ket{10}}{\sqrt{2}}.
\end{align}
\end{subequations}
In the Bell basis, a quoin-pair can be expressed as
\begin{equation}
\label{eq:p2}
\begin{aligned}
\ket{p}^{\otimes 2}
&= (\sqrt{1-p} \ket{0} + \sqrt{p} \ket{1})
   (\sqrt{1-p} \ket{0} + \sqrt{p} \ket{1}) \\
&= \dfrac{1}{\sqrt{2}}\ket{\Phi^+}
 + \dfrac{1-2p}{\sqrt{2}}\ket{\Phi^-}
 + \sqrt{2p(1-p)}\ket{\Psi^+}.
\end{aligned}
\end{equation}
It is clear from \eqref{eq:p2} that the probability $P(\Phi^+)$ of observing $\ket{\Phi^+}$ is $\frac{1}{2}$, enabling an efficient way of generating fair coins. We also find that
\begin{equation}
\label{eq:p_union}
    P(\Phi^+ \cup \Psi^-) = P(\Phi^- \cup \Psi^+) = \dfrac{1}{2},
\end{equation}
where $A \cup B$ represents observing $\ket{A}$ or $\ket{B}$. Thus, one fair coin can be generated using one joint measurement of two $p$-quoins. The corresponding quantum circuit is depicted in Fig.~\ref{fig:fig2}(a). This method is complementary to the one using one quoin and two measurements described earlier.

Now, we turn to functions that cannot be generated using CBF. These functions must violate one or more of the three conditions developed by Keane and O'Brien~\cite{Keane1994Bernoulli}: (i) $f(p)$ must be continuous, (ii) $f(p)$ must not approach a value of 0 or 1 exponentially quickly near $p = 0$ or 1, and (iii) $f(p) \neq \{0,1\}$ for $p \in (0, 1)$.

In particular, we focus on the following two functions:
\begin{equation}
f_\frown(p) = 4p(1-p), \ p \in [0, 1]
\end{equation}
and the famous ``Bernoulli doubling'' function
\begin{equation}
f_\wedge(p) = 2p \equiv 
\begin{cases}
2p & \text{if } p \in [0, 1/2] \\
2(1 - p) & \text{if } p \in (1/2, 1]
\end{cases}
\end{equation}
both of which violate condition (iii) for $p=\frac{1}{2}$ and can be utilized as a primitive for other functions~\cite{Nacu2005fast}.

From Eq.~\eqref{eq:p2} we observe that $P(\Psi^+) = 2p(1-p)$. However, if we restrict ourselves to the cases of observing $\ket{\Psi^+}$ or $\ket{\Phi^-}$ only, using Eq.~\eqref{eq:p_union} we obtain the conditional probability
\begin{equation}
\label{eq:p_cap}
    P\left(\Psi^+ | (\Phi^- \cup \Psi^+)\right) = 4p(1-p).
\end{equation}
Notably, one $f_\frown(p)$ coin consumes only four $p$-quoins on average, giving the protocol a constant and bias-independent overhead. This is because half of the Bell-basis measurements from two quoins result in $\ket{\Phi^-}$ or $\ket{\Psi^+}$.

Similarly, one can observe that it is possible to obtain $f_\wedge(p)$ by taking the square root (see Appendix~\ref{app:sqrt_p}) of the conditional probability 
\begin{equation}
\label{eq:p_sq}
    P\left(\Phi^- | (\Phi^- \cup \Psi^+) \right) = (1-2p)^2,
\end{equation}
and switching 0 and 1. Note that this method is distinct from the proposal by Dale \textit{et al.}~\cite{dale2015provable} where the function is first rewritten as $f_\wedge(p) = 1-\sqrt{1-4p(1-p)}$, followed by a series expansion
\begin{multline}
\begin{aligned}
    f_\wedge(p) &= \sum_{k=1}^{k_{\text{max}}}
    \left[
    \begin{pmatrix}
    2k \\ k
    \end{pmatrix}
    \dfrac{1}{(2k-1)2^{2k}} \right] (4p(1-p))^k \\
    &= \sum_{k=1}^{k_{\text{max}}} q_k (f_\frown(p))^k.
\end{aligned}
\end{multline}
A previous demonstration~\cite{Patel2019expt} using photonic systems implicitly used classical coins to generate an index $k$ with a probability $q_k$. It is, of course, possible to determine $q_k$ using fair coins generated in the Bell measurements. In contrast to using an $f_\frown(p)$ coin, we perform $f(p)=\sqrt{p}$ of Eq.~\eqref{eq:p_sq} completely using the fair coins~\cite{sqrt_p1999} generated from the Bell measurements (see Eq.~\eqref{eq:p_union}) for the Bernoulli doubling.

\section{Experimental results}

\begin{figure}[tb]
    \centering
    \includegraphics[width=\columnwidth]{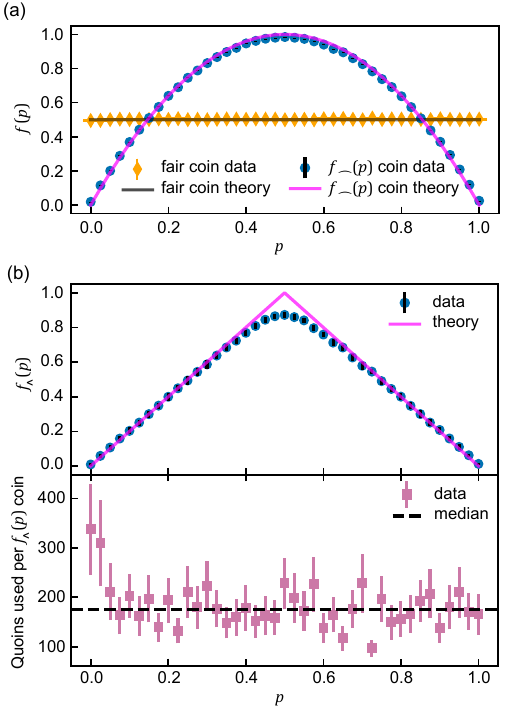}
    \caption{\textbf{Experimental quantum Bernoulli factory.} (a) Experimental data of a fair coin (orange diamonds) and a $f_\frown(p)$ coin (blue circles) as a function of bias $p$. The solid lines show ideal values. Each fair and $f_\frown(p)$ coin needs, on average, two and four quoins, respectively. Error bars are smaller than the marker size. (b) Top panel shows experimental data for the Bernoulli doubling function $f_\wedge(p)$ (blue circles) and the ideal values (magenta lines). The numbers of quoins used are shown as purple squares in the bottom plot with a median value of 176.1. Vertical error bars represent one standard deviation.}
    \label{fig:fig3}
\end{figure}

We perform the experiments using IBM's superconducting quantum platform on the device named `ibm\_sherbrooke'. The `Eagle r3' processor (see Appendix~\ref{app:ibm}) has 127 fixed-frequency superconducting transmon qubits connected in the `heavy-hexagon' geometry. We picked two adjacent qubits with $\mathtt{CNOT}$ error $7.0\times 10^{-3}$ and readout errors $\sim 1.0\times 10^{-2}$ while single-qubit gate errors are $<3.4\times 10^{-4}$. The qubits are initialized to the ground states $\ket{0}$ and the quantum circuit in Fig.~\ref{fig:fig2}(a) is applied, which results in the following mapping of the measurement basis to the Bell basis
\begin{equation}
\begin{split}
    \ket{00} &\rightarrow \ket{\Phi^+},\\
    \ket{01} &\rightarrow \ket{\Phi^-},\\
    \ket{10} &\rightarrow \ket{\Psi^+},\\
    \ket{11} &\rightarrow \ket{\Psi^-}.
\end{split}
\end{equation}
We sweep $p$ from 0 to 1, and for each value, $5\times10^4$ shots are recorded. The resulting measurement counts are plotted in Fig.~\ref{fig:fig2}(b). We process these data to generate fair coins as well as $f_\frown(p)$ and $f_\wedge(p)$ coins. In order to improve statistics and compute the standard deviation, we utilize bootstrapping where 200 sets are generated for every $p$, each containing $5\times10^4$ measurement entries. A fair coin is generated from each set using Eq.~\eqref{eq:p_union}, where two $p$-quoins are consumed. A heads outcome is output if the measurement is $\ket{\Phi^-}$ or $\ket{\Psi^+}$. Otherwise, the outcome is a tails outcome. The results are shown as orange diamonds with error bars in Fig.~\ref{fig:fig3}(a) which show excellent agreement with the theoretical values of $\frac{1}{2}$.

Similarly, generating $f_\frown(p)$ is straightforward where a heads (tails) outcome is assigned whenever the measurement yields $\ket{\Psi^+}$ $(\ket{\Phi^-})$ conditioned on not observing $\ket{\Phi^+}$ and $\ket{\Psi^-}$. The results are shown as blue circles in Fig.~\ref{fig:fig3}(a) which again show an excellent match with the theoretical prediction (magenta line). Error bars are plotted as one standard deviation and are smaller than the marker sizes in Fig.~\ref{fig:fig3}(a).

For the generation of $f_\wedge(p)$, we count the cases of obtaining $\ket{\Phi^-}$ conditioned on not observing $\ket{\Phi^+}$ and $\ket{\Psi^-}$ which results in an $f(p)=(1-2p)^2$ coin. Taking the square root of this function is done using a CBF protocol~\cite{sqrt_p1999} (see Appendix~\ref{app:sqrt_p}) where fair coins are needed. We utilize the fair coins generated by the same measurement without additional classical or quantum coins. Note that we do not reuse any consumed coin during the analysis. The results are shown as blue circles in Fig.~\ref{fig:fig3}(b) (top panel) which again show a very good match with the theoretical prediction (magenta line). The maximum value we obtain $\max[f_\wedge(p)]=0.873 \pm 0.016.$ The bottom panel shows the average number of quoins used to generate one $f_\wedge(p)$ as a function of $p$ where the median value is 176.1. This is calculated by dividing $10^5$ (consumed quoins per set) with the average number of $f_\wedge(p)$ coins generated. See Appendix~\ref{app:sqrt_p} for details. Note that this larger consumption is occurring due to the inefficiency of the CBF procedure for generating a $\sqrt{p}$-coin. The deviation from the theory around $p=0.5$ can be explained by the noise in the quantum hardware, dominated by the readout error (see Appendix~\ref{app:rdt_err}).

\section{Conclusions}

We have experimentally demonstrated a compact quantum Bernoulli factory based on Bell-basis measurements of two identical input quoins prepared on a superconducting-qubit processor. From a single joint measurement primitive we obtain, without relying on any external classical randomness source, an exact fair coin as well as the classically inconstructible function $f_\frown(p)=4p(1-p)$, each with constant average quoin cost (two and four, respectively) independent of the unknown bias $p$. Building on the same measurement statistics, we further implement the Bernoulli-doubling primitive $f_\wedge(p)$ and benchmark the observed output biases against ideal predictions.

Our data highlight both the practical viability and current limitations of entanglement-assisted randomness processing on noisy intermediate-scale quantum~\cite{Preskill2018NISQ} hardware. In particular, the dominant deviations near $p=1/2$ are consistent with readout misassignment, which sets an effective ceiling on the achievable $f_\wedge(p)$ in our implementation. These observations indicate clear experimental levers for improving performance, including measurement-error mitigation, improved calibration, and the use of qubits with lower readout and two-qubit gate errors.

More broadly, Bell-measurement-based primitives provide an experimentally simple route to generating multiple Bernoulli-factory coins from the same set of samples, and the ability to internally supply fair coins for subsequent classical subroutines (such as $\sqrt{p}$ constructions) removes an often implicit dependence on auxiliary randomness. Extending these ideas to larger entangled measurements, adaptive protocols, and integrated error-mitigation techniques may enable more general quantum-enhanced stochastic simulation and exact sampling primitives on near-term devices.


\section*{Acknowledgments}

This work was supported by the U.S. Department of Energy, Office of Science, National Quantum Information Science Research Centers, Superconducting Quantum Materials and Systems Center (SQMS), under Contract No. 89243024CSC000002. Fermilab is operated by Fermi Forward Discovery Group, LLC under Contract No. 89243024CSC000002 with the U.S. Department of Energy, Office of Science, Office of High Energy Physics. We also acknowledge the use of IBM Quantum services for this work.

\section*{Author Declarations}
\textbf{Conflict of Interest:} The author has no conflicts to disclose.

\section*{DATA AVAILABILITY}
The data that support the findings of this study are available from the corresponding author upon reasonable request.


\appendix 

\section{Generation of a $\sqrt{p}$-coin}
\label{app:sqrt_p}

One can use a CBF protocol as described in Ref.~\onlinecite{sqrt_p1999}. The Taylor expansion of $f(p)=\sqrt{p}$ around 1 takes the form
\begin{equation}
    \sqrt{p} = [1-(1-p)]^{1/2} = 1 - \sum_{n=0}^{\infty} \dfrac{C_n}{2^{2n+1}}(1-p)^{n+1},
\end{equation}
where $C_n = \begin{pmatrix} 2n \\ n \end{pmatrix}/(n+1)$ is the $n$-th Catalan number. $C_n$ can be defined as the number of possible configurations in a string containing $n$ ones and $n$ zeros such that the number of ones is equal or larger than the number of zeros in every initial segment. We use this property for generating $f(p)=\sqrt{p}$ coin. We continue to flip a fair coin until for the first time the number of heads is greater than that of the tails. If this happens after consuming $2n+1$ fair coins, we flip the $p$-coin $n+1$ times. The output is a tails if all coins are tails.

\begin{figure}[b]
    \centering
    \includegraphics[width=\columnwidth]{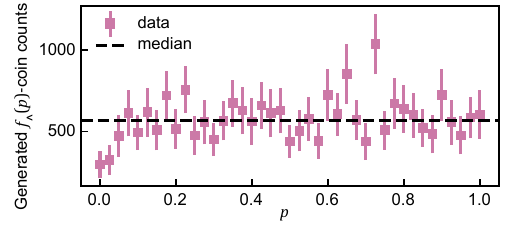}
    \caption{\textbf{Statistics of generated $f_\wedge(p)$ coin counts.} $10^5$ quoins are utilized for each $p$. The median value is 566.}
    \vspace{0.5em}
    \textbf{Supplemental Figure S1 Alt Text:} Scatter plot of generated Bernoulli doubling coin counts versus input bias p when 100 thousand quoins are used for each p. Purple square markers with vertical error bars show the generated coin counts; most values cluster around 500 to 650 counts with a few higher points near 900 to 1200 counts. A black dashed horizontal line marks the median count of 566.
    \label{fig:count}
\end{figure}

Importantly, the algorithm terminates with probability 1. This means the set of mathematical sequences that go on forever has a total probability of zero. One will not get stuck in an infinite loop forever in a real-world execution. The protocol, therefore, can use from one to a very large number of fair coins. Since, we have a finite number of fair coins (50,000 as 2 quoins generate one fair coin), in order limit runaway cases, we choose a maximum threshold of 5,000 fair coin consumptions after which those coins are discarded and the counter is reset. Figure~\ref{fig:count} shows the mean number of $f_\wedge(p)$ coins $n_x$ generated through this protocol where error bars $\sigma_x$ show one standard deviation.

Since one fair coin is generated using two quoins total number of quoins utilized is $10^5$. The mean number of quoins used per $f_\wedge(p)$-coin is calculated as $n_y=\frac{10^5}{n_x}$. Following, the standard ways of error propagation, the standard deviation $\sigma_y$ of $n_y$ is calculated as $\sigma_y = \frac{10^5}{n_x^2}\sigma_x=\frac{n_y}{n_x}\sigma_x$.

\section{IBM hardware}
\label{app:ibm}

\begin{table}[b]
\centering
\begin{tabular}{|c|c|}
\hline
Parameter & Value \\
\hline
Median T$_1$ & 394.77 $\mu$s \\
Median T$_2$ & 167.32 $\mu$s \\
Median ECR error & 7.477e-3 \\
Median readout error & 1.000e-2 \\
\hline
\end{tabular}
\caption{Various parameters of the `ibm\_sherbrooke' processor.}
\vspace{0.5em}
\textbf{Supplemental Table I Alt Text:}
Table listing representative hardware parameters for the IBM Sherbrooke processor. The table reports T1 of 394.77 microseconds, T2 of 167.32 microseconds, ECR two-qubit gate error of 7.477 parts per thousand and readout error of 0.01 as median values.
\label{tab:ibm}
\end{table}

We used one of the superconducting qubit chip named `ibm\_sherbrooke' provided by IBM's cloud based quantum service. It consists of 127 fixed-frequency transmon qubits connected in the heavy-hexagon geometry. The native gates used by the processor include ideal (ID), virtual rotation of about $z$-axis (RZ), $\pi$ rotation of the $x$-axis (X) and $\pi/2$ rotation about the $x$-axis (SX) as single qubit operations and echoed cross-resonance (ECR) as two-qubit entangling operations. Some relevant parameters are provide in Table~\ref{tab:ibm}. The runtime for the whole experiment was 8m 59s on the hardware which used qubit numbers \#113 and \#114.

\section{Readout error}
\label{app:rdt_err}

State misassignment during the readout plays an important role in limiting the performance of QBF. Let us consider that the correct assignment probabilities for the 0 and 1 states of the first qubit are $P(0|0)=a_0$ and $P(1|1)=a_1$ so that the same for false assignments are $P(1|0)=1-a_0 = \bar{a}_0$ and $P(0|1)=1-a_1=\bar{a}_1$. We use $b_j$ to represent the similar quantities for the second qubit. Then the two-qubit assignment or confusion matrix can be expressed as
\begin{equation}
    M = \begin{bmatrix}
        a_0 b_0 & a_0\bar{b}_1 & \bar{a}_1b_0 & \bar{a}_1\bar{b}_1\\
        a_0\bar{b}_0 & a_0 b_1 & \bar{a}_1\bar{b}_0 & \bar{a}_1b_1 \\
        \bar{a}_0b_0 & \bar{a}_0\bar{b}_1 & a_1b_0 & a_1\bar{b}_1 \\
        \bar{a}_0\bar{b}_0 & \bar{a}_0b_1 & a_1\bar{b}_0 & a_1b_1
    \end{bmatrix}.
\end{equation}
Note that the sum of each column equals 1. Let us calculate the effect of this measurement error for the case of Bernoulli doubling.  For $p=\frac{1}{2}$, the ideal final state is expected to be $\frac{\ket{00}+\ket{10}}{\sqrt{2}}$ so that the probability $P(01)$ of observing $\ket{01}$ is zero. However, the experimentally observed probability will be non-zero and can be computed as $\vec{P}'=M\times \vec{P}$ where $\vec{P}'=[P'(00) \ P'(01) \ P'(10) \ P'(11)]^T$ are the experimentally expected probabilities and $\vec{P}=[\frac{1}{2} \ 0 \ \frac{1}{2} \ 0]^T$ are the ideal probabilities. Let us consider $a_0=b_0=0.995$ and $a_1=b_1=0.985$ as ballpark numbers corresponding to the reported readout error of $10^{-2}$ (see Table~\ref{tab:ibm}) resulting in $P'(01)=0.0075=(1-2p)^2/2$ so that $2p=0.8775$ limiting the maximum value of $f_\wedge(p)$.

\bibliography{main}

\end{document}